\documentclass[aps,prb,twocolumn,showpacs]{revtex4}

\usepackage{bm}
\usepackage{color}
\usepackage[dvips]{graphicx}

\begin{document}

\title{
Ferromagnetic insulating phase in Pr$_{1-x}$Ca$_{x}$MnO$_3$ \\
}

\date{\today}

\author{R. Kajimoto}
\altaffiliation[Present address: ]{Neutron Science Facility,
Institute of Materials Structure Science, High Energy Accelerator
Research Organization, 1-1 Oho, Tsukuba, Ibaraki 305-0801, Japan}
\email{kaji@post.kek.jp.}
\affiliation{Department of Physics, Ochanomizu
University, Bunkyo-ku, Tokyo 112-8610, Japan}

\author{H. Mochizuki}
\affiliation{Neutron Science Laboratory, Institute for Solid State
Physics, University of Tokyo, Tokai, Ibaraki 319-1106, Japan}

\author{H. Yoshizawa}
\affiliation{Neutron Science Laboratory, Institute for Solid State
Physics, University of Tokyo, Tokai, Ibaraki 319-1106, Japan}

\author{S. Okamoto}
\altaffiliation[Present address: ]
{Department of Physics, Columbia University, 538 West 120th Street, New
York, NY 10027. }
\affiliation{Institute of Physical and Chemical Research (RIKEN),
Saitama 351-0198, Japan}

\author{S. Ishihara}
\affiliation{Department of Physics, Tohoku University, Sendai 980-8578,
Japan}

\begin{abstract}
 A ferromagnetic insulating (FM-I) state in
 Pr$_{0.75}$Ca$_{0.25}$MnO$_{3}$ has been studied by neutron scattering
 experiment and theoretical calculation. The insulating behavior is
 robust against an external magnetic field, and is ascribed to neither
 the phase separation between a ferromagnetic metallic (FM-M) phase and
 a non-ferromagnetic insulating one, nor the charge ordering. We found
 that the Jahn-Teller type lattice distortion is much weaker than
 PrMnO$_{3}$ and the magnetic interaction is almost isotropic. These
 features resembles the ferromagnetic metallic state of manganites, but
 the spin exchange interaction $J$ is much reduced compared to the FM-M
 state.  The theoretical calculation based on the staggered type orbital
 order well reproduces several features of the spin and orbital state in
 the FM-I phase.
\end{abstract}

% 75.25.+z Spin arrangements in magnetically ordered materials
% (including neutron and spin-polarized electron studies,
%  synchrotron-source x-ray scattering, etc.)
% 75.30.Ds Spin waves
% 75.47.Gk Colossal magnetoresistance
% 75.47.Lx Manganites

\maketitle

\section{Introduction}

Since the discovery of the high-$T_c$ superconductivity in layered
cupurates, doped Mott insulators with perovskite structures become one
of the major targets in recent studies of condensed matter physics. Hole
doped perovskite manganites, $R_{1-x}A_{x}$MnO$_{3}$ ($R$ is a trivalent
rare-earth ion and $A$ is a divalent alkaline-earth ion), have been
extensively studied since the discovery of the colossal
magnetoresistance (CMR) effect. Recent studies devoted to clarify an
origin of the CMR effect have revealed that the physics of manganites is
controlled by an interplay between spin, charge and orbital 
degrees of freedom.

At $x=0$ in $R_{1-x}A_{x}$MnO$_{3}$, a Mn ion shows an electron
configuration of $t_{2g}^{3}e_{g}^{1}$ where the $e_g$ orbital degree of
freedom is active. Actually, the mother compound $R$MnO$_{3}$ shows an
alternate ordering of the $d_{3x^2-r^2}$ and $d_{3y^2-r^2}$ orbitals
associated with the cooperative Jahn-Teller distortion in the $ab$ plane
of the orthorhombic \textit{Pbnm} lattice. The layer-type
(\textit{A}-type) antiferromagnetic (AF) spin ordering is realized at
low temperature by the anisotropic superexchange interaction under the
orbital ordering (OO).\cite{rodriguez98,murakami98,review} This
anisotropy was experimentally confirmed in LaMnO$_{3}$ by neutron
scattering studies.\cite{hirota96,moussa96} By substituting $R$ by $A$
in $R$MnO$_{3}$, holes are introduced in the system. In the simple
double exchange scenario, a kinetic energy gain of the doped holes
brings about the ferromagnetic metallic (FM-M) ground
state.\cite{review,gennes60} It is also supposed that the OO is
destroyed and the orbital liquid/disordered states are realized in the
FM-M phase.\cite{ishihara97b} The isotropic characters in the crystal
lattice\cite{urushibara95} and the spin exchange
interactions\cite{martin96,perring96} realized in this phase may be
attributed to this orbital state.

However, in the actual compounds, the doping dependence of the
electronic ground state is not so na\"{i}ve. Between the \textit{A}-type
AF phase and the FM-M phase, the ground state is ferromagnetic but
insulating.\cite{review} The ferromagnetic insulating (FM-I) state can
not be explained by the simple double exchange picture. Although such a
contradiction remains to be solved, the FM-I phase has not been much
studied so far. This paper provides a detailed information about the
character of the FM-I phase by a neutron scattering study and a
theoretical calculation on Pr$_{1-x}$Ca$_{x}$MnO$_{3}$, which is a
typical example for a FM-I phase with a relatively wide hole
concentration region $0.15 < x < 0.3$.\cite{tomioka96}

One of the interpretations of the FM-I phase is based on a phase
separation between the FM-M phase and a non-FM insulating phase. In a
recent trend of the research of CMR manganites, the phase separation
phenomenon is regarded as an important factor for the large resistivity
change accompanied by the insulator-metal (I-M) transition. A plausible
origin of the non-FM insulating phase is the \textit{CE}-type charge
ordering (CO).  This is a checkerboard-type CO of Mn$^{3+}$ and
Mn$^{4+}$ accompanied by a zig-zag ordering of $d_{3x^2-r^2}$ and
$d_{3y^2-r^2}$ orbitals on Mn$^{3+}$ sites and the \textit{CE}-type AF
spin ordering. The \textit{CE}-type CO is most stabilized at $x=1/2$,
but it is exceedingly robust against hole doping. Actually, in
Pr$_{1-x}$Ca$_{x}$MnO$_{3}$, the \textit{CE}-type CO is well developed
for $x \ge 0.3$ adjacent to the FM-I phase, and coexists with the FM
component.\cite{tomioka96,yoshi95,yoshi96} Recently, Dai \textit{et al.}
showed by neutron scattering measurement that the short-range order of
the \textit{CE}-type CO exists in the FM-I phase of
La$_{1-x}$Ca$_{x}$MnO$_{3}$,\cite{dai00} though the detail relation
between these charge correlations and the transport is not obvious.

The phase separation picture requires substantial volume fraction of
non-FM insulating phase to make the compound an insulator overcoming the
metallic conduction in the FM phase. Uehara \textit{et al.} studied the
phase separation between the FM-M phase and the charge-ordered
insulating (CO-I) phase in La$_{5/8-y}$Pr$_{y}$Ca$_{3/8}$MnO$_{3}$. They
showed that the concentration of the FM-M phase at the I-M transition,
which was evaluated from the ratio of the saturation magnetic moment to
the full moment of a Mn ion, agrees with the three-dimensional (3D)
percolation threshold 10--25\%.\cite{uehara99} However, the FM moment of
$\sim$2 $\mu_{B}$ in
Pr$_{0.7}$Ca$_{0.3}$MnO$_{3}$\cite{yoshi95,yoshi96,cox98} is too large
to account for its insulating character. On the other hand, there is
another possibility that the insulating nature does not arise from the
non-FM phase but it is intrinsic to the FM phase. Recently several kinds
of COs related to the hole concentration of $x=1/4$ were theoretically
proposed.  Hotta \textit{et al.} claimed that
Pr$_{3/4}$Ca$_{1/4}$MnO$_{3}$ becomes a FM-I state by a CO consisting of
a checkerboard charge ordered plane and a purely Mn$^{3+}$ plane
stacking alternately along the $z$ axis.\cite{hotta00} As a result, the
unit cell of the CO becomes $\sqrt{2} \times \sqrt{2} \times 4$ in terms
of the cubic perovskite unit (We will refer this type of CO as
$(\pi,\pi,\pi/2)$-type CO.).  Mizokawa \textit{et al.} proposed that in
the lightly doped manganites there exist ``orbital polarons.'' Here
Mn$^{4+}$ ions are surrounded by ferromagnetically coupled nearest
neighbor Mn$^{3+}$ ions where the $e_g$ orbitals are pointed to the
central Mn$^{4+}$ site.\cite{mizokawa00-1} They claimed that in
Pr$_{3/4}$Ca$_{1/4}$MnO$_{3}$, the orbital polarons form a body-centered
cubic lattice whose unit cell has a size of $2 \times 2 \times 2$ in
unit of the cubic perovskite lattice, and a FM insulator is
realized.\cite{mizokawa00-2} Neither kind of CO, however, has not been
detected experimentally so far. Instead, Endoh \textit{et al.}  claimed
by resonant x-ray scattering (RXS) and neutron scattering studies that
the FM-I phase of La$_{0.88}$Sr$_{0.12}$MnO$_{3}$ shows an OO of a
mixture of $d_{3z^2-r^2}$ and $d_{x^2-y^2}$ orbitals. They also proposed
that this OO is an origin of the FM-I state.\cite{endoh99,okamoto00} As
for the orbital state of the FM-I phase in Pr$_{1-x}$Ca$_{x}$MnO$_{3}$,
Zimmermann \textit{et al.}  advocated by a RXS study that the same OO as
that in LaMnO$_{3}$ (LaMnO$_{3}$-type OO) develops below about 500 K in
Pr$_{0.75}$Ca$_{0.25}$MnO$_{3}$.\cite{zimmermann01} In contrast to the
FM spin ordering in Pr$_{0.75}$Ca$_{0.25}$MnO$_{3}$, however,
LaMnO$_{3}$-type OO is generally accompanied with the \textit{A}-type AF
spin ordering.

The complicated status of the current research on the FM-I state of
manganites reminds us the necessity of a comprehensive study on spin,
charge, and orbital states of the FM-I phase of manganites. For this
purpose, neutron scattering is a useful tool to study CO and OO because
of its sensitivity to the lattice distortions induced by CO and OO. In
the present study, we performed elastic neutron scattering measurements
on a single crystal of Pr$_{0.75}$Ca$_{0.25}$MnO$_{3}$ to elucidate the
superlattice peaks and the lattice distortions due to CO and OO. The OO
affects the exchange interactions between spins due to the anisotropic
character of the $e_g$ orbitals. Therefore, through the estimation of
the exchange interactions by measuring the magnon dispersion relation,
we can characterize the orbital state. So, we also performed inelastic
neutron scattering measurements to determine the spin exchange through
the measurements of the spin wave excitations. 
The theoretical analyses provide a unified picture 
for the spin and orbital states in the AF, FM-I and FM-M phases. 

The rest of this paper is organized as follows.  In the next section the
experimental procedure is described.  The results and discussions are
described in Sec.\ III, where the property of the magnetic and crystal
structures in Pr$_{0.75}$Ca$_{0.25}$MnO$_{3}$ are discussed. In Sec.\
III-A, we will argue the possibility of the phase separation, and in
Sec.\ III-B we will characterize the FM-I state by measurements of the
structural properties and spin fluctuations. 
Section III-C is devoted to the theoretical investigation for the magnetic
interaction and the orbital state in moderately doped manganites. 
A summary of the work is given in Sec.\ V.

\section{Experimental procedures}

Single crystals of Pr$_{0.75}$Ca$_{0.25}$MnO$_{3}$ and PrMnO$_{3}$ were
grown by the floating zone method. A stoichiometric mixture of
Pr$_{6}$O$_{11}$, CaCO$_{3}$, and Mn$_{3}$O$_{4}$ was ground and
calcined at 1300{\char'27\kern-.35em\hbox{C}} in air for 12 h. Then the
resulting powder was pressed into rods and fired again at
1300{\char'27\kern-.35em\hbox{C}} in air for 12 h. The crystal growth
was performed in a floating-zone furnace in a flow of oxygen gas. We
precharacterized the samples by a powder x-ray diffraction measurement
and confirmed that the samples are single phase. Then the samples were
cut into the size of $6\,\mbox{mm}\phi \times 30\,\mbox{mm}$ for the
present neutron scattering study. The transport data were taken using
Quantum Design PPMS on a sample cut from the same batch as used in the
neutron study.

The neutron scattering experiments were performed using triple axis
spectrometers GPTAS and HQR of Institute for Solid State Physics,
University of Tokyo, installed at the JRR-3M research reactor in JAERI,
Tokai, Japan.  The most of the measurements were done at GPTAS, while
HQR was utilized when a high $Q$ resolution was required. At both
spectrometers, the 002 reflection of pyrolytic graphite (PG) was used
for the monochromator and analyzer. At GPTAS, we selected a neutron wave
length of $k_f = 3.81$ {\AA}$^{-1}$. For elastic scattering
measurements, a combination of horizontal collimators of
20$^{\prime}$-40$^{\prime}$-40$^{\prime}$-80$^{\prime}$ (from reactor to
detector) was adopted and two PG filters were placed before the
monochromator and after the sample to suppress contaminations of
higher-order harmonics. For inelastic scattering measurements, we chose
40$^{\prime}$-40$^{\prime}$-40$^{\prime}$-40$^{\prime}$ combination of
collimators and set a PG filter after the sample. At HQR, which is on
the thermal guide tube, neutrons of wave length $k_i = 2.56$
{\AA}$^{-1}$ and horizontal collimation of
blank-40$^{\prime}$-80$^{\prime}$ (from monochromator to sample) were
used together with a PG filter before the sample. The crystals were
mounted in aluminum cans filled with helium gas. The temperature of the
samples were controlled by conventional and high-temperature-type closed
cycle He-gas refrigerators. The measurements were performed in the
$(h,0,l)$ and $(h,h,l)$ zones of reciprocal space of the \textit{Pbnm}
lattice.

\section{Results and discussions}

\subsection{Possibility of the phase separation}

%\subsubsection{Order parameter and transport property}

\begin{figure}
 \centering
 \includegraphics[scale=0.6]{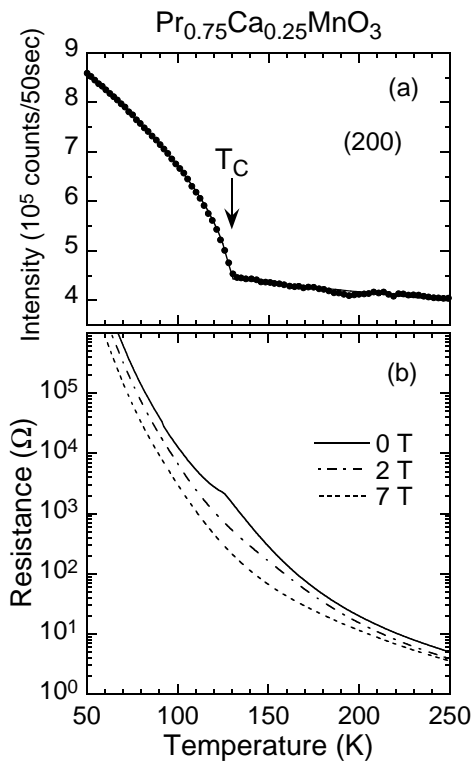}
 \caption{(a) Temperature dependence of the scattering intensity of the
 200 peak. (b) Temperature dependence of the resistance under the
 magnetic field of $H=0$ T, 2 T, and 7 T.}
 \label{resist}
\end{figure}

First, let us characterize the sample by showing the results of the
temperature ($T$) dependences of the order parameter of the magnetic
order and the resistance. Figure \ref{resist}(a) is the $T$ dependence
of the scattering intensity of the fundamental 200 peak. The intensity
starts to increase below the Curie temperature $T_C \sim 130$ K due to
the FM spin ordering, consistent with a previous study.\cite{tomioka96}
To determine the critical exponent, we fit the data to a power law,
$(1-T/T_C)^{2\beta}$, varying the lower limit of the temperature range
of the fitting, $T_\mathrm{lim}$. $\beta$ was determined to be 0.36(1)
as $T_\mathrm{lim} \to T_C = 129.5(5)$ K.  This value of the critical
exponent is very close to that for the 3D FM Heisenberg model, 0.367,
and similar to that for the FM-M manganites such as
La$_{0.7}$Sr$_{0.2}$MnO$_{3}$\cite{doloc98} and
La$_{0.7}$Sr$_{0.3}$MnO$_{3}$.\cite{martin96,doloc98} The magnetic
moment was determined from a powder sample which was prepared by
crushing a melt-grown crystal. The FM moment is 3.32(3) $\mu_B$ per Mn
site at 15 K, oriented parallel to the $b$ axis. This value of the FM
moment is nearly equal to the formula value of 3.75 $\mu_B$.  This means
that even if a non-FM region existed in the sample, its volume would be
extremely small.

Figure \ref{resist}(b) shows the $T$ dependence of the resistance at
several magnetic fields. For $H=0$ T, the sample is insulating for all
temperature range measured, and a small cusp is observed at $T_C$. By
applying the magnetic field, this cusp vanishes, and the resistance is
reduced from that at $H=0$ T. However the insulating behavior still
persists under the magnetic field up to $H=7$ T. The robustness of the
insulating state contrasts markedly with the \textit{CE}-type charge
ordered phase for $x>0.3$, where the CO melts by a magnetic field of a
few teslas.\cite{tomioka96} In the present case, because the spins are
already fully aligned ferromagnetically at $H=0$ T, an external field
cannot make a significant change of the transports.

%\subsubsection{CE-type CO}

\begin{figure}
 \centering
 \includegraphics[scale=0.5]{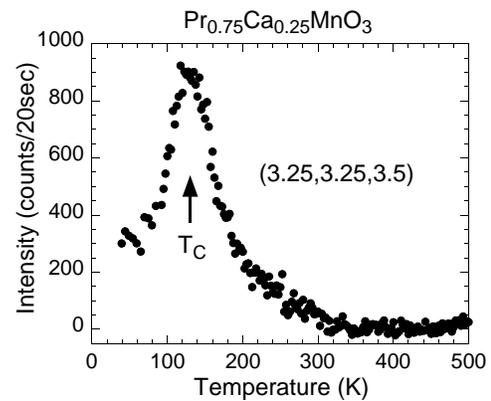}
 \caption{Temperature dependence of the intensity of the CE-type
 charge/orbital order peak at $\bm{Q}=(3.275,3.275,3.55)$. The intensity
 due to the Huang scattering measured at $\bm{Q}=(3.33,3.33,3.66)$ is
 subtracted.}
 \label{CE}
\end{figure}

As mentioned above, the magnetic and transport properties suggest that
the insulating behavior of Pr$_{0.75}$Ca$_{0.25}$MnO$_{3}$ does not
originate from the non-FM region, but it is rather intrinsic to the FM
state. To confirm this point, we performed the neutron diffraction
measurements and surveyed reciprocal points to detect signals due to the
\textit{CE}-type CO, and found the lattice superlattice peaks due to the
\textit{CE}-type CO. Figure \ref{CE} shows a $T$ dependence of the
scattering intensity at $\bm{Q}=(3.25,3.25,3.5)$, which corresponds to a
\textit{CE}-type CO peak $(5,1.5,0)$\cite{comm} of a different domain of
the sample. Because diffuse scattering due to single polarons (Huang
scattering) overlaps at this position,\cite{shimomura99} the intensity
of the diffuse scattering estimated at $\bm{Q}=(3.33,3.33,3.66)$ is
subtracted from the data, and intensity of 100 counts is added to
compensate the difference in background intensity at the two
positions. The \textit{CE}-type CO develops below $T \sim 320$ K. It is
drastically suppressed below $T_C$, although the sample remains
insulating [Fig.~\ref{resist}(b)].  The small cusp of the resistance at
$T_C$ may be related to the melting of the \textit{CE}-type CO, but the
overall insulating feature of the resistance, however, cannot be
explained by the \textit{CE}-type CO. This is in contrast to the FM-M
region of manganites, where the disappearance of the short-range
\textit{CE}-type CO accompanies the transition to the FM-M
state.\cite{shimomura99,dai00,adams00} We think that the existence of
the \textit{CE}-type CO is due to a small mixture of a higher Ca
concentration phase. We also found the pseudo \textit{CE}-type AF peaks
concomitant with the \textit{CE}-type CO in the region of $x <
1/2$.\cite{yoshi95,yoshi96,cox98,jirak85} Their intensity is of the
order of $10^{-3}$ in comparison with that of the FM peak, indicating
that the volume of the \textit{CE}-type CO phase is negligible.

\subsection{Character of the homogeneous FM-I state}

\begin{figure}
 \centering
 \includegraphics[scale=0.5]{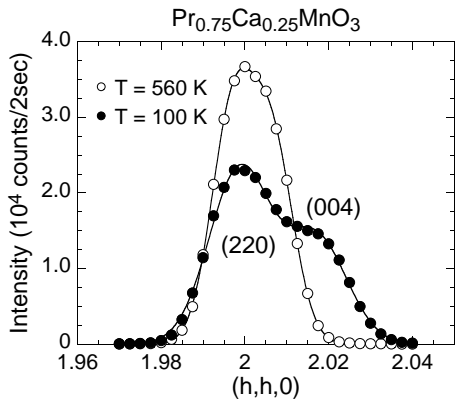}
 \caption{Profiles of the 220/004 doublet along the [110] direction
 at 100 K and 560 K. The data are collected at the HQR
 spectrometer. Solid lines are fits to two Gaussians.}
 \label{220prof}
\end{figure}

%\subsubsection{Possibility of charge ordering}

%\subsubsection{Dagotto's CO}

The results described in the previous section indicates the phase
separation between the FM-M phase and the non-FM insulating phase is not
important for the FM-I phase of Pr$_{1-x}$Ca$_{x}$MnO$_{3}$. The FM-I
state is an intrinsic state of the low doped region of
manganites. Consequently, to understand the \textit{homogeneous} FM-I
phase, we will characterize the charge and orbital states in this phase.

First, we tried to clarify whether the FM-I state is a FM charge ordered
state as theoretically proposed. For the $(\pi,\pi,\pi/2)$-type CO
proposed by Hotta \textit{et al.},\cite{hotta00} the superlattice peaks
are expected at $\bm{Q}=(h,k,l)$ with $h+k=\mbox{odd}$ and $l=\mbox{half
integer}$. To detect this type of CO, we surveyed $(m,0,1/2)$ and
$(5,0,n/2)$ positions where $m$ and $n$ are odd integer, but could not
find any peaks. This result is negative for the existence of this type
of charge order in Pr$_{0.75}$Ca$_{0.25}$MnO$_{3}$.

%\subsubsection{Mizokawa's OO}

The orbital polaron lattice proposed by Mizokawa \textit{et
al.}\cite{mizokawa00-1,mizokawa00-2} will produce superlattice
reflections at $\bm{Q}=(h,k,l)$ with ($h+k=\mbox{odd}$ and
$l=\mbox{even}$) or ($h,k=\mbox{half integer}$ and $l=\mbox{odd}$). We
found peaks at several positions satisfying these conditions and
forbidden by the \textit{Pbnm} symmetry. Unfortunately, however, these
positions coincide with those with integer indices considering the
twinning of the crystal. Because additional small tilts of MnO$_{6}$
octahedra can produce Bragg peaks at such positions, we cannot determine
whether the orbital polaron lattice exists or not from the present
study, though no clear relation between the intensities of these peaks
and the resistance was observed.

%\subsubsection{Orbital ordering}

Next, we will investigate the orbital state of the FM-I state. According
to the recent RXS study, resonant signals observed in
Pr$_{0.75}$Ca$_{0.25}$MnO$_{3}$ is consistent to the same OO as
LaMnO$_{3}$, i.e. alternate ordering of $d_{3x^2-r^2}$ and
$d_{3y^2-r^2}$ orbitals in the $ab$ plane.\cite{zimmermann01}
LaMnO$_{3}$-type OO is expected to produce significant anisotropies in
crystal structure as well as spin exchange: The in-plane lattice
constants $a$ and $b$ become longer than the out-of-plane lattice
constant $c$, and the exchange interactions within the $ab$ plane will
be stronger than that along the $c$ axis. Therefore, measuring these
quantities should be useful for quantitative understanding of the
orbital state.

%\subsubsection{Crystal Structure}

\begin{figure}
 \centering
 \includegraphics[scale=0.5]{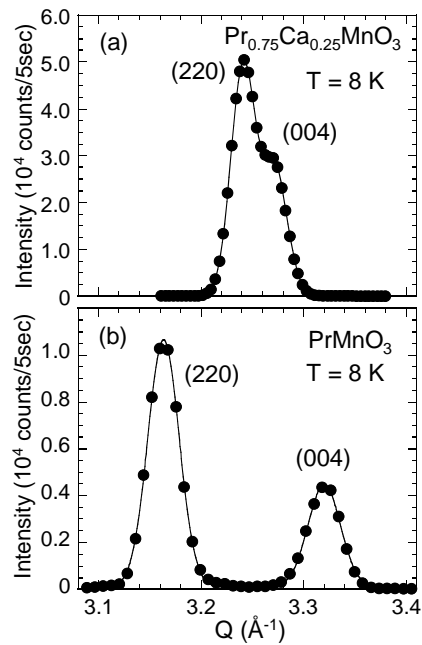}
 \caption{Profiles of the 220/004 doublets along the [110]
 direction for (a) Pr$_{0.75}$Ca$_{0.25}$MnO$_{3}$ and (b) for
 PrMnO$_{3}$ at 8 K. The horizontal axes indicate the amplitudes of the
 scattering vector $Q$. Solid lines are fits to two Gaussians.}
 \label{220compare}
\end{figure}

Figure \ref{220prof} shows neutron diffraction profiles around the
fundamental (2,2,0) position along the [110] direction at 100 K and 560
K. Because of the twinning, the 004 peak is also observed at slightly
higher $Q$ than the 220 peak. At $T=560$ K, both peaks locate so close
that the observed profile forms almost a single peak. On the other hand,
this peak splits into two peaks at $T=100$ K due to small increase of
the $a$ or $b$ axis and large decrease of the $c$ axis. This change of
the lattice constants is consistent to the cooperative Jahn-Teller
distortions accompanied by the LaMnO$_{3}$-type OO. We measured $T$
dependences of similar profiles and found that the structural change
gradually develops below $\sim$450 K. This temperature coincides with
the temperature where the orbital signals measured by the RXS study
starts to increase,\cite{zimmermann01} and corresponds to the transition
temperature from the $O$ phase to the $O^{\prime}$ phase identified in
Ref.\ \onlinecite{jirak85}. These results qualitatively suggest the
developing of the LaMnO$_{3}$-type OO below $T \sim 450$ K. However, the
amplitude of the lattice distortion is fairly small compared to the
mother compound, PrMnO$_{3}$, where a typical LaMnO$_{3}$-type OO is
expected. This can easily be assured by comparing the profile of the
220/004 reflections for both compounds.  In Fig.~\ref{220compare}, we
depicted the profiles of the 220 and 004 reflections of
Pr$_{0.75}$Ca$_{0.25}$MnO$_{3}$ [Fig.~\ref{220compare}(a)] and
PrMnO$_{3}$ [Fig.~\ref{220compare}(b)] at $T = 8$ K. We show the
amplitude of the scattering vector $Q$ as the horizontal axis for easier
comparison of two data.

%\subsubsection{Structural analysis}

%\subsubsection{Spin wave}

\begin{figure}
 \centering
 \includegraphics[scale=0.45]{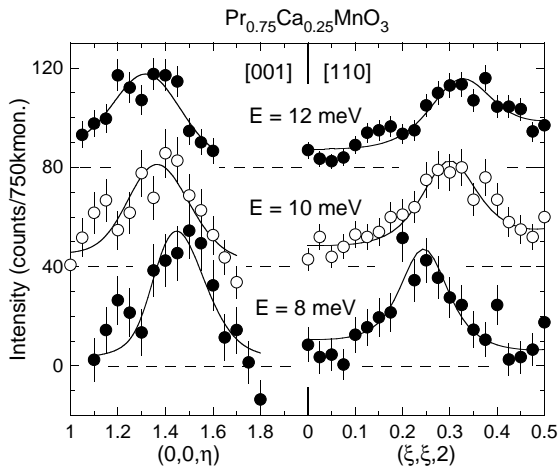}
 \caption{Profiles of constant-energy scan of the magnon spectra along
 the [001] direction (left) and the [110] direction (right). The
 background is subtracted from each data. The vertical axes are shifted
 by 40 counts for each energy. Solid lines are fits to the Lorentzian
 spectral function convoluted with the instrumental resolution.}
 \label{sw_prof}
\end{figure}

\begin{figure}
 \centering
 \includegraphics[scale=0.5]{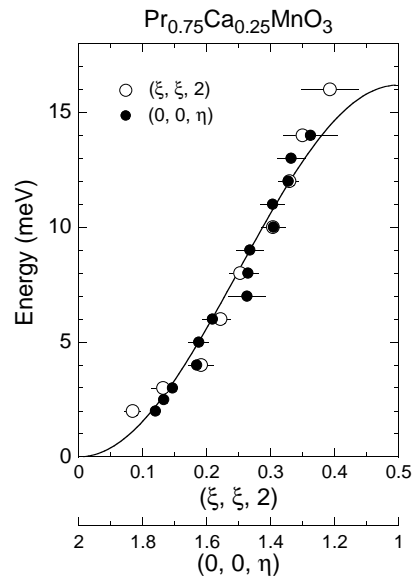}
 \caption{Spin wave dispersion relations in
 Pr$_{0.75}$Ca$_{0.25}$MnO$_{3}$ along the [110] and 
 [001] directions at 8 K. Solid line is a dispersion curve for the
 FM Heisenberg model with an isotropic nearest neighbor coupling.}
 \label{sw_disp}
\end{figure}

Figure~\ref{sw_prof} is typical profiles of constant-energy scans of
spin wave excitations along the out-of-plane direction [001] (left
panel) and the in-plane direction [110] (right panel) of
Pr$_{0.75}$Ca$_{0.25}$MnO$_{3}$. The data were collected near the FM
Bragg point (0,0,2) at $T=8$ K. The magnetic origin of these excitations
were confirmed by measuring their $Q$ dependence. The peak positions get
away from the zone center as the energy increases according to the
dispersion relation of the spin wave. From these measurements, we
obtained the dispersion relation of the spin wave along the two
directions, which is shown in Fig.~\ref{sw_disp}. Note that the momentum
transfer $\bm{q} = (0,0,2n)$ has almost the same magnitude as $\bm{q} =
(n,n,0)$ in the \textit{Pbnm} notation. In contrast to the
LaMnO$_{3}$,\cite{hirota96,moussa96} two dispersion curves well
coincide, indicating that the spin exchange is isotropic ($J_{ab} \sim
J_{c}$). This is consistent to the critical exponent $\beta$ [Sec.\
III-A], which is almost the same as the value for a 3D system. The
observed dispersion curves can be described by the conventional FM
Heisenberg model: $\mathcal{H} = - \sum_{ij} J_{ij}
\bm{S}_{i}\cdot\bm{S}_{j}$, where $J_{ij}$ is the exchange integral
between spins at site $\bm{R}_{i}$ and $\bm{R}_{j}$. In the linear
approximation, the dispersion relation is given by $\hbar\omega(\bm{q})
= \Delta + 2S[J(\bm{0})-J(\bm{q})]$, where $J(\bm{q}) =
\sum_{j}J_{ij}\exp[i\bm{q}\cdot(\bm{R}_{i}-\bm{R}_{j})]$ and $\Delta$
accounts for small anisotropies. Taking only the isotropic nearest
neighbor coupling $J$ into account is sufficient to reproduce the
observed data, and we get the solid line in Fig.~\ref{sw_disp} with $2JS
= 4.0(2)$ meV and $\Delta = 0.0(3)$ meV.

The isotropic character of the spin system shows an interesting
resemblance among the orbital-ordered FM-I phase and the
orbital-disordered FM-M phase. However, one can distinguish them by the
magnitude of the spin exchange.  Clear enhancement of the spin stiffness
$D(T)$ on the transition from the FM-I phase to the FM-M phase has been
observed in a hole concentration dependence of
La$_{1-x}$Sr$_{x}$MnO$_{3}$\cite{endoh97} and
La$_{1-x}$Ca$_{x}$MnO$_{3}$\cite{dai01} by neutron scattering studies.
The same phenomenon on the field-induced I-M transition in
Pr$_{0.7}$Ca$_{0.3}$MnO$_{3}$ has also been observed.\cite{baca02} In
these materials, $D(0)$ in the insulating phase is $\approx$50
meV\AA$^{2}$ and that in the metallic phase is $\approx$150
meV\AA$^{2}$. If we estimate $D(0)$ of the present
Pr$_{0.75}$Ca$_{0.25}$MnO$_{3}$ compound from $J$ at 8 K by a relation
$D=2J S a_{c}^2$ where $a_{c}$ is the lattice constant of the cubic
lattice, $D(0)$ becomes $\approx$60 meV\AA$^{2}$. This value is almost
same as that of the insulating La$_{1-x}$Sr$_{x}$MnO$_{3}$,
La$_{1-x}$Ca$_{x}$MnO$_{3}$, and Pr$_{0.7}$Ca$_{0.3}$MnO$_{3}$. This
fact evidences the common ground of the FM-I phase of the manganites.

The Jahn-Teller type lattice distortion means that there may exist a
LaMnO$_{3}$-like OO as reported by the RXS study. Its smallness and the
isotropy in $J$, however, suggest that the electron distribution is more
isotropic than that in the LaMnO$_{3}$-type OO.  Similar situation is
realized in the FM-I phase of
La$_{0.88}$Sr$_{0.12}$MnO$_{3}$,\cite{hirota97,endoh99} where the
staggered orbital ordering of
$(|d_{3z^2-r^2}\rangle\pm|d_{x^2-y^2}\rangle)/\sqrt{2}$ is
proposed.\cite{endoh99,okamoto00} In the next section, we will
theoretically investigate the orbital state Pr$_{1-x}$Ca$_{x}$MnO$_{3}$
and construct the spin and orbital phase diagram.

\subsection{Theory}

Spin and orbital states in manganites with moderate hole doping are
examined theoretically from the view point of strong electron
correlation.  We adopt the $t$-$J$ type Hamiltonian where the double
degeneracy of the $e_g$ orbitals is taken into account.  The model
Hamiltonian consists of the following four terms: \cite{ishihara97}
\begin{equation}
 {\cal H} = {\cal H}_{t}+{\cal H}_{J}+{\cal H}_{H}+{\cal H}_{AF}  . 
  \label{eq:ham}
\end{equation}
The first and second terms are the main terms corresponding to the $t$
and $J$ terms in the $t$-$J$ model, respectively, given by
\begin{equation}
 {\cal H}_t = \sum_{\langle i j \rangle \gamma \gamma' \sigma}
 t_{ij}^{\gamma \gamma'} {\widetilde d}_{i \gamma \sigma}^\dagger {\widetilde d}_{j \gamma' \sigma} 
 +H.c. , 
\end{equation}
and 
\begin{eqnarray}
 {\cal H}_{J}
  &=& -2J_1 \sum_{\langle ij \rangle} 
       \left ( 
	\frac{3}{4} n_i n_j+\vec S_i \cdot \vec S_j
       \right )
       \left (
	\frac{1}{4}-\tau_i^l \tau_j^l
       \right ) \nonumber \\
  &-& 2J_2 \sum_{ \langle ij \rangle }
       \left (
	\frac{1}{4} n_i n_j -\vec S_i \cdot \vec S_j
       \right )
       \left (
        \frac{3}{4} + \tau_i^l \tau_j^l+\tau_i^l+\tau_j^l
       \right ) . 
\end{eqnarray}
${\widetilde d}_{i \gamma \sigma}$ is the annihilation operator of the
$e_g$ electron at site $i$ with orbital $\gamma$ and spin $\sigma$.
This operator is defined in the restricted Hilbert space where the $e_g$
electron numbers at each Mn site are limited to be one or less.  The
prefactor $t_{ij}^{\gamma \gamma'}$ is the transfer integral between
orbital $\gamma$ at site $i$ and $\gamma'$ at $j$.  ${\cal H}_{J}$
describes the superexchange interactions between the nearest neighboring
Mn spins and orbitals.  The spin and orbital degrees of freedom for the
$e_g$ electron are represented by the spin operator
\begin{equation}
 \vec S_i =
  \frac{1}{2}
  \sum_{\gamma \sigma \sigma'} 
  {\widetilde d}_{i \gamma \sigma}^\dagger 
  \vec \sigma_{\sigma \sigma'} 
  {\widetilde d}_{i \gamma \sigma'} , 
\end{equation}
and the pseudo-spin operator
\begin{equation}
 \vec T_i =
  \frac{1}{2}
  \sum_{\gamma \gamma' \sigma} 
  {\widetilde d}_{i \gamma \sigma}^\dagger 
  \vec \sigma_{\gamma \gamma'} 
  {\widetilde d}_{i \gamma' \sigma} , 
\end{equation}
respectively, with the Pauli matrices $\vec \sigma$.  $\tau_i^l$ in
${\cal H}_J$ is defined by the pseudo-spin operator as
\begin{equation}
 \tau_i^l = \cos \left ( \frac{2 m_l \pi}{3} \right ) T_{i z}
  +\sin \left ( \frac{2 m_l \pi}{3} \right ) T_{i x} , 
\end{equation}
with $(m_x, m_y, m_z)=(1,2,3)$ where $l(=x,y,z)$ indicates a direction
of the bond connecting site $i$ and site $j$.  The magnitudes of the
superexchange interactions $J_1$ and $J_2$ satisfy the condition
$J_1>J_2>0$.  The third and fourth terms in the Hamiltonian
Eq.~(\ref{eq:ham}) describe the Hund coupling $J_H(>0)$ between the
$e_g$ and $t_{2g}$ spins
\begin{equation}
 {\cal H}_H = -J_H \sum_{i} \vec S_i \cdot \vec S_{t i } , 
\end{equation}
and the AF superexchange interaction $J_{AF}(>0)$ between the nearest
neighboring $t_{2g}$ spins
\begin{equation}
 {\cal H}_{AF} = J_{AF} \sum_{\langle ij \rangle} 
  \vec S_{ti} \cdot \vec S_{t j} , 
\end{equation}
respectively.  $\vec S_{t i}$ is the spin operator for the $t_{2g}$
electrons with $S=3/2$.  It was confirmed that this model is successful
in description the spin and orbital ordered state in hole doped and
undoped manganites.\cite{ishihara97,okamoto00}

Here, we examine the orbital ordered and disordered states in the
unified fashion by applying the generalized slave boson method.  The
electron annihilation operator ${\widetilde d}_{i \gamma \sigma}$ is
decomposed into the holon $h_i$, spinon $s_{i \sigma}$ and pseudo-spinon
$t_{i \gamma}$ operators as
\begin{equation}
 {\widetilde d}_{i \gamma \sigma} = h_i^\dagger s_{i \sigma} t_{i \gamma} ,  
\end{equation}
with the local constraints of 
\begin{equation}
 h_i^\dagger h_i+\sum_\gamma t_{i \gamma}^\dagger t_{i \gamma} = 1, 
\end{equation}
and 
\begin{equation}
 \sum_\sigma s_{i \sigma}^\dagger s_{i \sigma}
  = \sum_{\gamma} t_{i \gamma}^\dagger t_{i \gamma}, 
\end{equation}
at each Mn site.  We choose that both the holon and spinon are bosons,
and the pseudo-spinon is a fermion. \cite{khaliullin00,oles02,hirota02}
This is because 1) in the
ferromagnetic metallic phase, the hole carrier is recognized to be a
fermion with orbital degree of freedom which is treated by $h_i^\dagger
t_{i \gamma}$, and 2) the fermionic operator for the orbital degree of
freedom $t_{i \gamma}$ is suitable for describing the orbital disordered
state.  We adopt the mean field approximation where we take the bond
order parameters for orbital
\begin{equation}
 \sum_{\gamma \gamma'} \left \langle  t_{i i+l}^{\gamma \gamma'} 
  t_{i \gamma}^\dagger t_{i+l \gamma'} \right \rangle = t_0 \chi_t^l, 
\label{eq:chit}
\end{equation}
and those for spin 
\begin{equation}
 \sum_{\sigma } \left \langle  s_{i \sigma}^\dagger s_{i+l \sigma}
  \right \rangle = \chi_s^l , 
\end{equation}
and the site diagonal order parameters for orbital  
\begin{equation}
 \sum_{\gamma \gamma'}
  \left \langle  t_{i \gamma}^\dagger \sigma_{\gamma \gamma'}^\mu t_{i \gamma} 
  \right \rangle=m_{t}^{\mu}(i) , 
\end{equation}
for $\mu=x$ and $z$, and those for spin 
\begin{equation}
 \sum_{\sigma } \left \langle
                 s_{i \sigma}^\dagger \sigma_{\sigma \sigma}^z s_{i \sigma} 
                \right \rangle = m_{s}^z(i) . 
\end{equation}
$t_0$ in Eq.~(\ref{eq:chit}) is the electron transfer intensity between
the nearest neighboring $d_{3z^2-r^2}$ orbitals along the $z$ axis.  We
also introduce a mean field for the holon
\begin{equation}
 \left \langle h_i^\dagger h_i \right \rangle
  = - \left \langle h_i^\dagger h_j \right \rangle
  = x ,
\end{equation} 
with the hole concentration $x$.  In this scheme, the local constraints
are released into the global ones:
\begin{equation}
 \sum_i \left (
         h_i ^\dagger h_i+\sum_\gamma t_{i \gamma}^\dagger t_{i \gamma}
        \right ) = 1 , 
\end{equation}
and 
\begin{equation}
 \sum_{i \sigma} s_{i \sigma}^\dagger s_{i \sigma}
  = \sum_{i \gamma} t_{i \gamma}^\dagger t_{i \gamma} . 
\end{equation}  
By considering the strong Hund coupling $J_H$, the directions of $\vec
S_i$ and $\vec S_{ti}$ are assumed to be the same.  The effects of the
lattice distortion and its coupling to the $e_g$ electrons are not
considered explicitly in the present model.  This is because, in the
moderately hole doped region of our interest, the Jahn-Teller distortion
is significantly reduced as shown in the previous section.

\begin{figure} 
%\epsfxsize=0.9\columnwidth 
%\centerline{\epsffile{fig1.eps}} 
 \centering
 \includegraphics[width=0.9\columnwidth,clip]{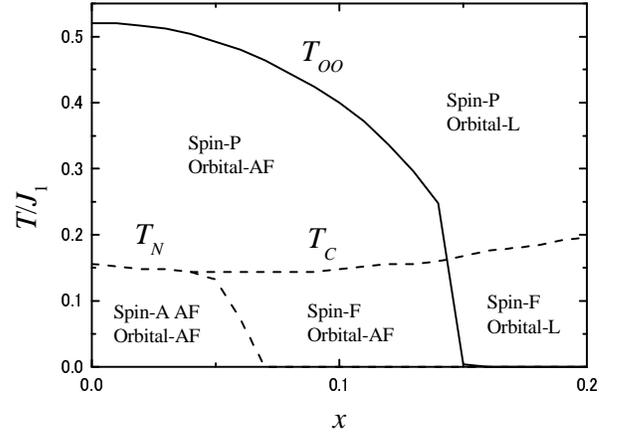}
 \caption{The spin and orbital phase diagram in the hole concentration
 $x$ and temperature $T$ plane.  The parameter values are chosen to be
 $J_2/J_1=0.25$, $J_{AF}/J_1=0.014$, and $t_0/J_1=2$.  $F$, $AF$, $P$
 and $L$ indicate the ferromagnetic phase, the antiferromagnetic phase,
 the paramagnetic phase, and the orbital liquid phase, respectively.}
 \label{fig:th1} 
\end{figure}

The phase diagram is obtained numerically by minimizing the free energy
with respect to the order parameters [Fig.~\ref{fig:th1}].  The
parameter values are chosen to be $J_2/J_1=$0.25, $J_{AF}/J_1$=0.014,
and $t_0/J_1=2$.  For the Fourier transform of the site diagonal order
parameters $m_t^\mu(\vec k)$ and $m_s^z(\vec k)$, we assume that $m_t^x
(\pi \pi 0)$, $m_t^z (000)$, $m_s^z (0 0 \pi)$ and $m_s^z(0 0 0 )$ are
finite in the ordered phases.  Below the orbital ordering temperature
$T_{OO}$, the site diagonal orbital order parameters $m^{\mu}_t(\vec k)$
appear.  $T_C$ and $T_N$ are the Curie temperature and the N\'{e}el
temperature for the \textit{A}-type AF order, respectively, where
$m_s^z(000)$ and $m^z_s(0 0 \pi )$ appear, respectively.  At $x=0$, the
staggered type orbital order $ m_t^x(\pi \pi 0)$ occurs far above $T_N$
as observed in LaMnO$_3$ and PrMnO$_3$.  By taking into account the
cooperative Jahn-Teller effects which is not included in the present
calculation, $T_{OO}$ may increase from the calculated value.  With
doping of holes, $T_{OO}$ rapidly decreases and disappears around
$x=0.15(\equiv x_c)$.  Above $x_c$, $m^{\mu}_t(\vec k)$ becomes zero,
but the bond order parameters $\chi_t^l(\vec k)$ still remain finite.
That is, the orbital liquid state is realized above the critical point
$x_c$.  As for the magnetic ordering, the $A$-type AF phase is changed
into the FM phase through the canted AF one.  The $x$ dependence of
$T_C$/$T_N$ is almost flat below $x=0.1$ and slightly increases with
increasing $x$.  The ground state electronic phase are classified into
(I) the (canted) $A$-type AF phase with the staggered OO corresponding
to LaMnO$_3$ and PrMnO$_3$, (II) the FM phase with the staggered OO
which is discussed later in more detail, and (III) the FM phase with the
orbital liquid state corresponding to the FM-M state in heavily hole
doped manganites, such as La$_{1-x}$Sr$_x$MnO$_3$ with $x \sim 0.3$.  In
a series of Pr$_{1-x}$Ca$_x$MnO$_3$, it is interpreted that this phase
does not appear because the charge ordered AF insulating phase is stable
down to the region of $x \sim 0.3$ instead of this FM-M phase.

Now we focus on the FM phase with the staggered OO.  We interpret that
this corresponds to the FM phase in Pr$_{1-x}$Ca$_x$MnO$_3$ with
$0.15<x<0.3$ of our present interest.  The OO in this FM phase is of the
staggered type where the two kinds of orbital are characterized by the
orbital mixing angles $(\theta/-\theta)$, as well as the OO in the
$A$-type AF phase.  This angle is defined as $|\theta\rangle =
\cos\frac{\theta}{2}|d_{3z^2-r^2}\rangle +
\sin\frac{\theta}{2}|d_{x^2-y^2}\rangle$.  This is consistent with the
azimuthal angle scan of the RXS experiments in
Pr$_{0.75}$Ca$_{0.25}$MnO$_3$; the azimuthal angle $\varphi$ is the
rotation angle of the sample around the x-ray scattering vector $\vec
k_i-\vec k_f$ in the RXS experiments.  By measuring the $\varphi$
dependence of the RXS intensity $I(\varphi)$, the information for the
symmetry of the orbital ordered states is deduced.\cite{ishihara98} The
experimentally observed $I(\varphi)$ in Pr$_{0.75}$Ca$_{0.25}$MnO$_3$ is
proportional to a function $\sin^2 \varphi$ which is consistent with the
theoretical prediction for $I(\varphi)$ in the OO state of the
$(\theta/-\theta)$ type.\cite{zimmermann01,ishihara98} The value of
$\theta$ in the present calculation is about $\pi$.  That is, the
orbital wave functions are given by $(|d_{3z^2-r^2} \rangle \pm |
d_{x^2-y^2} \rangle)/\sqrt{2}$ where the electronic clouds are more
elongated along the $z$ direction in comparison with the $d_{3x^2-r^2}$
and $d_{3y^2-r^2}$ orbitals.  It is noted that, with doping of holes,
the saturated orbital moment $|m_t^{\mu}|$ gradually decreases, and
finally disappears at $x_c$.  On the other hand, the bond variable
$|\chi^{l}_t|$ grows up from zero at $x=0$.  That is, the two FM
interactions, i.e.  the double exchange interaction based on the kinetic
term ${\cal H}_t$ and the superexchange FM one from ${\cal H}_J$,
coexist.  This is consistent with the present experimental results in
Pr$_{0.75}$Ca$_{0.25}$MnO$_3$ that the magnitude of the Jahn-Teller
distortion, related to the magnitude of the diagonal order parameters
$m_t^\mu$, is significantly reduced in comparison with that in
PrMnO$_3$.  Unfortunately, in this calculation, this FM phase is not
insulating, because the translational symmetry is assumed in the mean
field approximation.  However, we numerically confirm that the band
width of charge carriers, which is proportional to the bond order
parameter $\chi_t^l$, is much reduced in the orbital ordered FM phase in
comparison with that in the FM phase orbital liquid phase.  We believe
that doped carriers in such a narrow band tend to be localized by other
ingredients which are not taken into account in the present model.

\begin{figure} 
%\epsfxsize=0.9\columnwidth 
%\centerline{\epsffile{fig2.eps}} 
 \centering
 \includegraphics[width=0.9\columnwidth,clip]{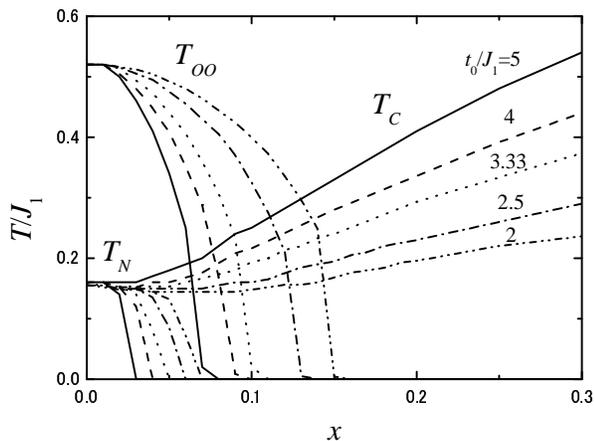} 
 \caption{The spin and orbital phase diagram as functions of $t_0/J_1$. 
 The parameter values are chosen to be $J_2/J_1=0.25$ and
 $J_{AF}/J_1=0.014$.} 
 \label{fig:th2} 
\end{figure} 
We also examine the phase diagram by changing the relative ratio of the
kinetic energy and the exchange energy $t_0/J_{1(2)}$
[Fig.~\ref{fig:th2}].  With decreasing $t_0/J_{1(2)}$, the critical hole
concentration $x_c$, where $T_{OO}$ disappears, increases and the region
of the orbital ordered FM phase is extended to the region with higher
$x$.  On the other hand, $T_C$ in the FM metallic phase is reduced and
$T_N$ and $T_C$ become less sensitive to $x$.  These calculated results
may explain a difference of the phase diagrams of
La$_{1-x}$Sr$_x$MnO$_3$ (LSMO), La$_{1-x}$Ca$_x$MnO$_3$ (LCMO) and
Pr$_{1-x}$Ca$_x$MnO$_3$ (PCMO).  The parameter value of $t_0/J_{1(2)}$
in the calculation is expected to decrease in this order of the
materials.  The FM-I phase is realized in the regions of $0.1<x<0.15$
for LSMO, $0.12<x<0.22$ for LCMO and $0.15<x<0.3$ for PCMO; the FM-I
phase is extended and shifts to the region with higher $x$.  $T_C$ in
the FM-M phase in LSMO is almost twice of $T_N$ in LaMnO$_3$.  On the
other hand, difference between $T_N$ in PrMnO$_3$ and $T_C$ in PCMO with
$x=0.3$ is about 30$\%$ of the $T_N$.

\begin{figure} 
%\epsfxsize=0.9\columnwidth 
%\centerline{\epsffile{fig3.eps}} 
 \centering
 \includegraphics[width=0.9\columnwidth,clip]{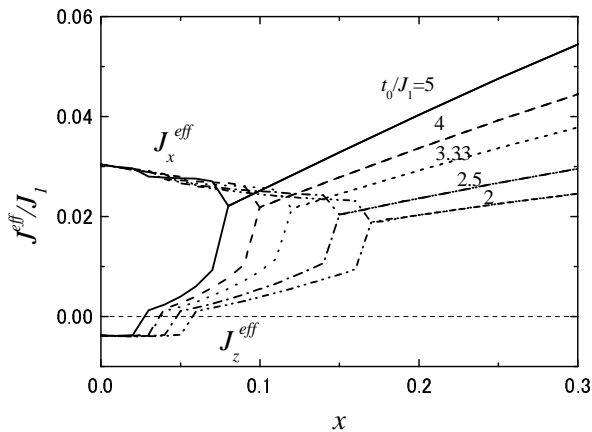}
 \caption{The $x$ dependence of the effective exchange interaction
 $J^\mathit{eff}_x$ and $J^\mathit{eff}_z$. The parameter values are
 chosen to be $J_2/J_1=0.25$ and $J_{AF}/J_1=0.014$.}
 \label{fig:th3} 
\end{figure}

In order to examine the spin dynamics in the orbital ordered FM phase,
we calculate the spin stiffness by applying the linear spin wave theory
to the model Hamiltonian in Eq.~(\ref{eq:ham}).  The effective exchange
interaction $J^\mathit{eff}$ corresponding to the interaction $J_l$
along the direction $l(=x,y,z)$ in the Heisenberg model $-\sum_{\langle
ij \rangle} J^\mathit{eff}_l \vec S_i \cdot \vec S_j$ is obtained as
\begin{equation}
 J^\mathit{eff}_l = \frac{-D^t_l+2D^J_l}{2S^2} . 
\end{equation}
$D^t_l$ is a stiffness constant arising from the transfer term of the
Hamiltonian ${\cal H}_t$ given by
\begin{equation}
 D_l^t = xt_0\chi_t^l , 
\end{equation}
and $D_l^J$ is the one from ${\cal H}_J$ and ${\cal H}_{AF}$ given by
\begin{equation}
 D_l^J = -J_l \frac{1}{4}-J_{AF} \frac{9}{4} , 
\end{equation}
with 
\begin{eqnarray}
 J_x &=& -\frac{1}{2}(J_1-3J_2)(1-x)^2 \nonumber \\
     &+& \frac{1}{2}(J_1+J_2)
     \left ( \frac{1}{4} m_t^{z 2}-\frac{3}{4} m_t^{x 2}-4 \chi_t^{x 2} \right)
     -J_2 m_t^z, 
\end{eqnarray}
and 
\begin{eqnarray}
 J_z &=& -\frac{1}{2}(J_1-3J_2)(1-x)^2 \nonumber \\
     &+& \frac{1}{2}(J_1+J_2) \left ( \frac{1}{2}m_t^{z 2}-4\chi^{z2}_t \right)
         +2 J_2 m_t^z . 
\end{eqnarray}
Here, we define the average magnitude of the Mn spin as $S=(1-x)/2+3/2$.
The $x$ dependence of the effective exchange interactions, $J_x(=J_y)$
and $J_z$, are presented in Fig.~\ref{fig:th3}.  In the $A$-type AF
phase near $x=0$, $J_z$ is negative and its magnitude is much weaker
than that of $|J_x|$.  This is experimentally confirmed by the inelastic
neutron scattering in LaMnO$_3$.  With doping of holes, the sign of
$J_z$ is changed into a positive one, and $J_x$ decreases and $J_z$
increases.  That is, the anisotropy of the exchange interactions is much
reduced.  This tendency of the $x$ dependence of the anisotropy of
$J^\mathit{eff}_l$ is consistent with the present neutron scattering
experiments in Pr$_{1-x}$Ca$_x$MnO$_3$, although, in the calculation,
the weak anisotropy of $J^\mathit{eff}$ remains in the orbital ordered
FM phase. Such an anisotropy may not be detected due to the coarse
resolution of the present neutron scattering experiments.

\section{Conclusion}

We have systematically investigated the nature of the FM-I state of
perovskite manganites by a neutron scattering study on a single crystal
of Pr$_{0.75}$Ca$_{0.25}$MnO$_{3}$ and theoretical calculations. We have
found that 1) the insulating behavior is robust against an external
magnetic field, 2) the FM moment is almost fully polarized, 3) the
Jahn-Teller lattice distortions are strongly suppressed, and 4) the spin
exchange interactions are isotropic. These features are against
interpretations of insulating behavior based on COs or phase separation,
and are consistent to the staggered type orbital ordered state.  The
spin stiffness is similar to those observed in other FM-I manganites and
has a smaller value than that in the FM-M phase, indicating the
ubiquitous origin of the FM-I state. We theoretically construct the
comprehensive spin and orbital phase diagram for the FM phase, which can
explain the difference of the phase diagrams of several manganites.

Lastly, we should note the importance of our work for the research of
the I-M transition or the CMR phenomenon in hole-doped perovskite
manganites. Many previous experimental researches focusing attention on
the phase separation as an origin of the I-M transition attribute an
isotropic phase or a FM phase to a metallic phase.  However, the
existence of an \textit{isotropic ferromagnetic} insulating phase means
one must pay better attention to identifying the metallic phase and
consider the effect of the staggered type OO especially in the low-doped
region.

\begin{acknowledgments}

This work was supported by a Grant-In-Aid for Scientific Research from
the Ministry of Education, Culture, Sports, Science, and Technology,
Japan. R. K. was supported by the Japan Society for the Promotion of
Science.

\end{acknowledgments}

\end{document}